# Variation in defects and properties in composite of ZnO and α-Fe$_2$O$_3$ for methyl blue dye removal

Boris Wareppam[1], K. Priyananda Singh[1], N. Joseph Singh[1], Subrata Ghosh[2,*], Ng. Aomoa[3], V.K. Garg[4], A. C. Oliveira[4], L. Herojit Singh[1,*]

[1] *Department of Physics, National Institute of Technology Manipur, Langol 795004, India*

[2] *Micro and Nanostructured Materials Laboratory – NanoLab, Department of Energy, Politecnico di Milano, via Ponzio 34/3, Milano - 20133, Italy*

[3] *Centre of Plasma Physics-Institute for Plasma Research, Sonapur 782402, Assam, India*

[4] *Institute of Physics, University of Brasília, 70919-970 Brasília, DF, Brazil*

**ABSTRACT**

The plasma deposition wall coated composite of ZnO and α-Fe$_2$O$_3$ (ZF-W) after exposure to ~ 2000 °C, mostly considered as waste-materials and cleaned out from the deposition unit, was subjected to anneal at 300, 500 and 1000 °C to manipulate the structural properties. An evolution of defect states along with the structural changes has been identified as annealing temperature was varied. As a consequence, an unstable state of ZnFe$_2$O$_4$ was found to be stabilized at 500 °C and migration of Zn from ZnO causes the phase transformation from the α-Fe$_2$O$_3$ to ZnFe$_2$O$_4$. While implemented for methyl blue adsorption/degradation without the effect of any external sources, the degradation for ZF-W annealed at 300 °C, 500 °C and 1000 °C were 84%, 68% and 82%, respectively. Compared to annealed structures, pristine ZF-W delivered the highest methyl blue adsorption efficiency of 86%. The changes in adsorption/degradation properties have been correlated with the simultaneous evolution of defects and structural properties of ZF-W as annealed at different temperatures. The plausible mechanism on the interaction of methyl blue with the composites on the adsorption/degradation is proposed. These findings give a clear indication on the importance of defects presence in the mixed metal oxide composite to obtain high-performance degradation/adsorption properties for sustainable wastewater treatment.

# I. INTRODUCTION

Structuring metal oxides at nanoscale have garnered a significant attention due to their intriguing properties and widespread applications [1–4]. The related applications includes their use in catalysis, photocatalysis, solar cells, gas sensors, batteries, magnetic storage media, optoelectronic, and electronics [5–8]. The performance of active materials for desired application rely on the geometry and structural properties, whereas the properties of a material depends on the synthesis method, compositions, cation distribution, size of the nano-crystallites, amount of defects etc. [9].

Amongst, cations have its own choice of distribution in the matrix depending on the environment and synthesis conditions. For instance, some cations have a strong preference for occupying a specific interstitial site [9]. In some cases, high annealing temperatures are much-needed to introduce the cations at their preferred (minimum energy) site. Whereas, in many other cases (at lower temperatures), they easily occupy a random site because of low thermal energy to cross the energy barrier to an ordered state [10–12]. It is noteworthy to mention, in particular for spinel structures, that annealing the samples at high temperature leads to reduction of the inverse spinel properties [13]. A metastable state is considered for transition of normal-inverse-normal spinel structure. Thus, it is understandable that the degree of inversion is heavily dependent on the processing time and the temperature, for example in Ball-milling method. The inverse to normal transition in some spinel start from 600- 800 °C [13]. That being said, cation distribution in some spinel is greatly affected even at low temperature. A few cations require high annealing temperature to achieve their stable state while others may not even be stable at high temperature. Hence, engineering the nanostructures with appropriate cation distribution is key to implement for desired applications.

Furthermore, cation inversion is heavily influenced by the presence of oxygen vacancy [14]. The occurrence of oxygen in oxides material plays another vital role in altering the intrinsic properties of nanocomposites [15]. Changes in oxygen stoichiometry, for example in spinel ferrites, can alter the super exchange path and give rise to rearrangement of ions. These can also give rise to various defects mainly cation and oxygen vacancies. Previous research has looked into the role of native defects in superior photocatalytic activity of ZnO and $CeO_2$ nanoparticles where the presence of oxygen vacancies narrows the band gap and accelerates the charge separation thereby regulating the electronic structure [16,17]. The importance of oxygen vacancies in facilitating adsorption and activation of reactant molecules have also been well established [18,19]. When oxygen vacancies come into contact with the reactants, they can modify their state (bond length, bond angle, coordination pattern, or intermediates), enhancing their attraction for the reactants [20].

Along with the oxygen vacancies, metal oxide nanostructures also have a diverse defect chemistry, including vacancies, interstitial or substitutional defects, anti-sites, structural distortion etc. [21]. The emergence of defects, their interrelations, and photocatalytic activity are all intrinsically complex and the understanding is not clear yet. Due to the large number of surface active sites and high surface energy, introducing intrinsic defects in materials has been proven to be a potential strategy for improving their performance [20,22,23]. The creation of a large number of structural and microstructural defects can sometimes result in numerous polymorphic phase transitions [24,25]. It is utmost important to analyze these defect states in order to understand the phase transformation. As such, the defect concentration of the materials can be modulated by annealing to obtain the controlled structural and optical properties. In the field of photo-catalysis, published literature data reveal that intrinsic defects like oxygen vacancies ($V_O$) and zinc interstitials ($Zn_i$) might improve the photocatalytic activity of produced ZnO nanomaterials against model pollutant species like dyes of various sorts [21]. It has also been revealed that these defect centres are temperature dependent. ZnO annealed at higher temperature (~ 550 °C) shows strong orange luminescent defects related to $V_O$-$Zn_i$ defect-complex [26]. Annealing of single phase metal oxides compound at high temperature in the atmospheric pressure leads to decrease in the amount of defects [27]. Similarly, annealing of single phase $ZnFe_2O_4$ nanoparticles up to 650 °C results to the decrease in the quadrupole splitting from ~ 0.48 to ~ 0.38 mm/s. This is understood as due to the reduction of point defects in the crystal after annealing [28]. However, the annealing effect of mixed phase composites require a thorough study, especially in the case of metal oxides where intra-particle cationic diffusion occurs which may generate point defects.

Designing materials may not be worthy unless it exhibits promising features in desired application. As a consequence of industry's rapid expansion, the modern globe is already experiencing a shortage of clean and pure water [29–31]. Color compounds, particularly dyes, are the most released contaminants among numerous wastes generated by industries with hazardous impact on both aquatic organisms and human life [32]. As a result, dye degradation, particularly photocatalytic dye degradation or adsorption, is of great interest. It has already been reported that the degradation of dye is influenced by morphology, surface binding sites, surface area and most importantly the native defects [33–36]. Dye adsorption of several anionic dye has been reported to be strongly dependent on the native defects [33,35]. ZnO is known to have good photocatalytic properties under ultra-violet (UV) irradiation while spinel ferrites are superb photocatalyst under both visible and UV light source. More importantly, spinel ferrite are highly abundant, have high adsorption properties and surface to volume ratio [37]. Harmful organic dye degradation under light irradiation have been frequently studied and expensive for large-scale productivity. Also, UV light source present in the natural sunlight is very low [38]. Here, the synergistic catalytic effect of ZnO and iron oxides made it possible for the degradation of dye under the dark condition.

Based on the above-mentioned background, the scope of this study is to explore the annealing effect of the already synthesized ball mill mixture of ZnO and α-$Fe_2O_3$ (1:1 molar %) which was exposed to high density DC plasma torch at a pressure of $10^{-3}$ mbar. Detailed synthesis procedure was reported in our previous work [39]. Herein, the sample ZF-W (collected from the wall of the chamber) constituting ZnO, α-$Fe_2O_3$ and cubic spinel ferrites are annealed at 300 °C, 500 °C and 1000 °C and their properties in comparison to ZF-W are discussed. The variation in the evolution of defect states and its role in the adsorption of Methyl Blue (MB) dye with the effect of annealing temperature are being discussed here.

## II. EXPERIMENTAL METHODS
### A. Materials

Methyl Blue (MB) was purchased from Himedia Pvt. Ltd. All reagents were of analytical grade and were used without further purification. Double distilled water was used throughout the experiment.

### B. Synthesis Method

Mixture of ZnO and $Fe_2O_3$ having stoichiometric molar ratio of 1:1 is treated in a plasma chamber as described in the our previous report [39]. The sample from the wall (ZF-W) is collected and annealed at different temperature of 300 °C, 500 °C and 1000 °C which are marked as ZF-W(300), ZF-W(500) and ZF-W(1000), respectively.

### C. Characterisation and measurement

Phase and crystallographic information of the plasma-treated powders were evaluated using X-ray diffraction (XRD) (Bruker-D8 advance-USA) with Cu $K_\alpha$ radiation of wavelength 1.54056 Å. $Fe^{57}$ Mössbauer spectroscopy was utilized in transmission mode with initial activity of 25 mCi (Wissel, Germany). The Photoluminescence (PL) emission spectra of composites were recorded using HITACHI F- 7000 Fluorescence spectrophotometer with a Xe source at an excitation wavelength of 280 nm at room temperature. The degradation of methyl blue (MB) was recorded using UV–Vis spectrometer (Shimadzu-1800-Japan) under continuous scan in the wavelength range of 350-800 nm.

### D. Degradation of Methyl Blue Dye

In the catalytic degradation of MB dye, the initial concentration of 50 ppm MB dye solution was prepared at room temperature. 1 mg of catalyst was added to 4 ml MB solution in Quartz curvette. The experiment was performed under static condition with continuous scan in dark condition without exposure to any light source. The initial pH check of MB dye solution was found to be 7.0 and pH value was not affected after adding the catalyst. The amount of degradation of MB due to the catalyst is calculated using the relation:

$$\% \ degradation = \left(1 - \frac{C}{C_0}\right) X \ 100$$

where $C_0$ is the initial concentration of the methyl blue, $C$ is the MB concentration at certain time $t$.

## III. Results and Discussion
### A. XRD analysis

The XRD spectra of pristine ZF-W, ZF-W(300) and ZF-W(500), as shown in Figure 1(a-b), contain combined phases comprising of ZnO, α-$Fe_2O_3$ and cubic spinel ferrites ($Fe_3O_4$ and $ZnFe_2O_4$). However, the sample ZF-W(1000) contain phases of ZnO and cubic spinel ferrites only (Figure 1(a-b)). The quantification of the phases was performed with the help of Match software (Table 1), which reveals that the content of hematite phase decreases as annealing temperature increased and it disappears completely at 1000 °C. This may be due to fact that hematite phase is transformed into cubic spinel ferrites as a result of Zn migration into the α-$Fe_2O_3$ lattices. Further clarification has done through analysis of Mossbauer spectra, which is discussed in next section. Additionally, the crystallite size of the cubic spinel ferrites was estimated using Scherrer's formula from the (220) diffraction plane. The crystallite size of ZF-W(300) decreases compare to ZF-W due the fact that the transition where coalescence of the crystallite starts above 300 °C. As reported, the phase transition of magnetite to hematite occurs typically within a temperature range of 275-375 °C [40,41]. Moreover, the decrease in the crystallite size of the cubic spinel phase at 300 °C may plausibly be due to oxidation of fraction of cubic spinel at the surface to the hematite phase. Beyond 300 °C, the crystallite size was found to increase as the annealing temperature increases (Figure 1(c)). This indicates that the small crystallites are merged with adjacent crystallites to form a larger crystallites [42] and the point defects are decreasing during the annealing process.

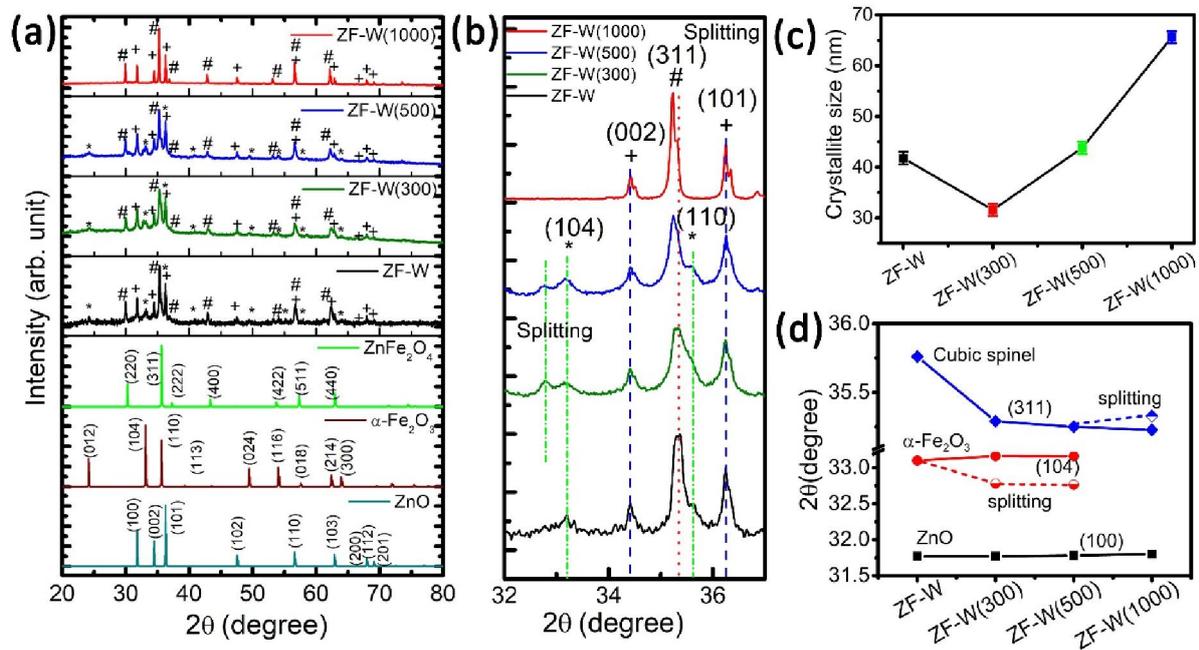

**Figure 1**: a) XRD patterns of ZF-W, ZF-W(300), ZF-W(500) and ZF-W(1000). The powder diffraction patterns of the component materials are provided for comparison. (Assigned) + →ZnO, ∗ →α-Fe$_2$O$_3$, # →cubic spinel (ZnFe$_2$O$_4$/ Fe$_3$O$_4$). (b) enlarge XRD patterns showing splitting of peaks; Variation of (c) crystallite size and (d) peak position against increase in annealing temperature.( The lines are only intended as guides for the eye)

Table 1. Quantification data of the phases of the prepared samples.

| Sample | Quantification (%) | | | Crystallite Size of Cubic spinel ferrites (in nm) |
|---|---|---|---|---|
| | ZnO | α-Fe$_2$O$_3$ | ZnFe$_2$O$_4$ | |
| **ZF-W** | 24 | 30 | 46 | 41.78 |
| **ZF-W(300)** | 27 | 27 | 46 | 31.59 |
| **ZF-W(500)** | 31 | 21 | 48 | 43.81 |
| **ZF-W(1000)** | 31 | - | 69 | 65.61 |

Figure 1(d). shows the shifting of peaks for cubic spinel phase while α-Fe$_2$O$_3$ and ZnO remains unchanged. These shifting can be correlated with the lattice parameter. The lattice constant of the cubic spinel was calculated for each plane (220), (222), (400), (422) and (511). In order to draw a comparison between the average lattice constant $a$ and true value of lattice constant $a_0$, the Nelson-Riley extrapolation function F(θ) was calculated for each reflection using the relation [43]:

$$F(\theta) = \frac{1}{2}\left[\frac{cos^2\theta}{sin\theta} + \frac{cos^2\theta}{\theta}\right]$$

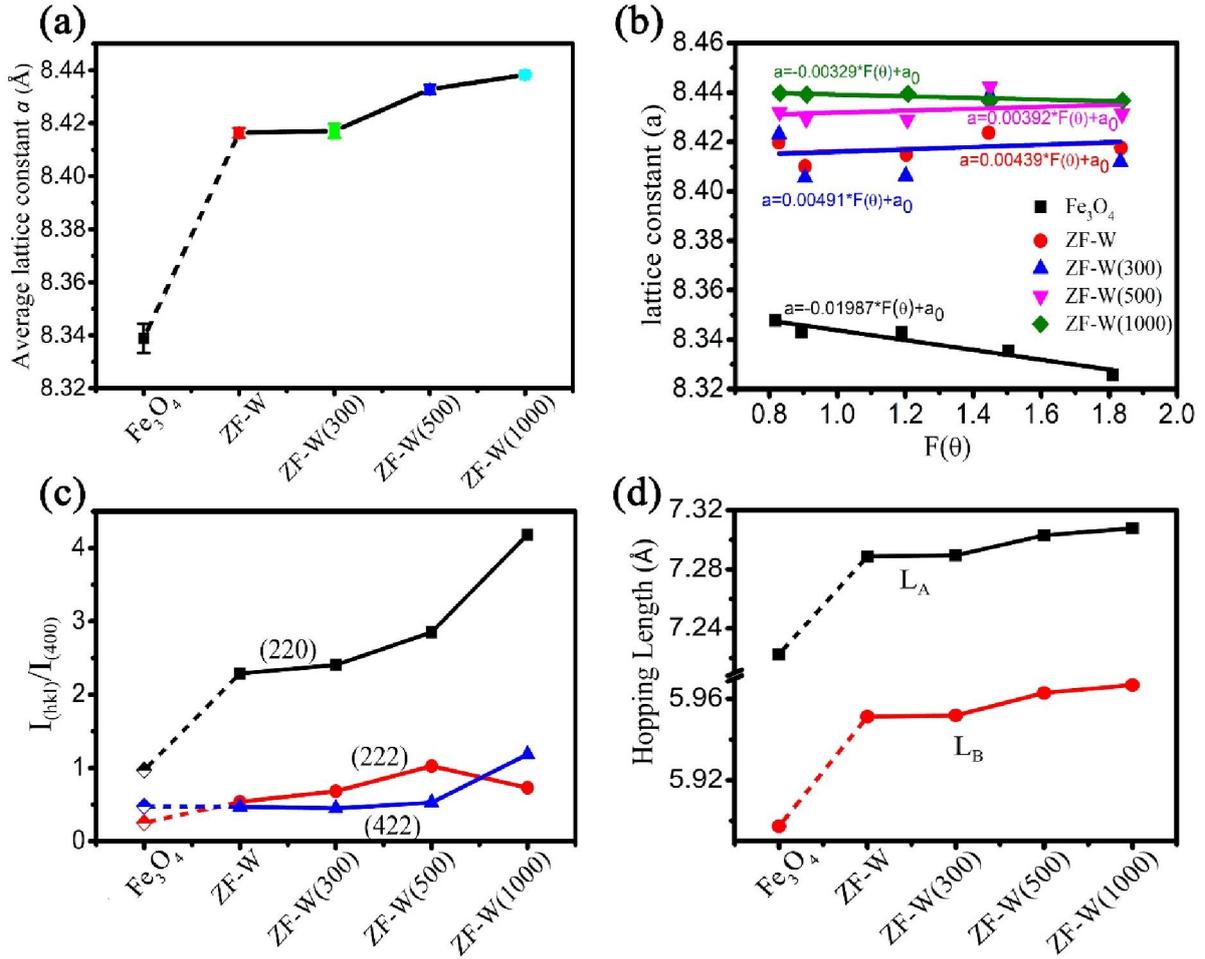

**Figure 2**: (a) Variation of lattice parameter *a* with the increase in annealing temperature and (b) as a function of $F(\theta)$; (c) Relative Intensity $I(hkl)/I(400)$ with varying annealing temperature and (d) the plot of hopping length $L_A$ and $L_B$ in A and B sites. The line is just a guide for the eye.[ data presented for $Fe_3O_4$ is taken from the previously reported work just for comparison [44,45]]

The plot of $F(\theta)$ *vs* lattice constant *a* for the samples is shown in Figure 2(b). The true value of the lattice constant $a_0$ can be extracted by extrapolating the line at $F(\theta)=0$. There is a slight variation in the values of the lattice parameters. It is noteworthy to mention that, in both the cases, the lattice constant increases as the annealing temperature increases (Figure 2(a-b)). This could be related with the cation distribution. As the annealing temperature increases, the cations migrate from one site to the other. Thus, the increase in lattice parameters could be due to the substitution of $Fe^{3+}$-ions in tetrahedral sites by the $Zn^{2+}$-ions as the temperature increases. Moreover, the intensity ratio $I_{(hkl)}/I_{(400)}$, which is reported to probe the cation distribution, [46] in cubic spinel phase were also estimated and shown in Figure 2(c). It is being reported that the intensity ratio corresponding to (220) and (422) diffraction planes are related to cations on tetrahedral sites while the (222) reflection corresponds to the octahedral site [46,47]. It is found that the intensity ratio increases as the annealing temperature increases in the tetrahedral sites. Interestingly, a dip is observed for the sample annealed above 500 °C in the octahedral sites. This fact can be explained on the basis of cation migration. The increase could be due to the substitution of smaller radius $Fe^{3+}$ ions (0.60Å) with a larger radius $Zn^{2+}$ ions(0.74Å) in the tetrahedral and octahedral sites [46]. Fraction of $Zn^{2+}$ ion enters both in tetrahedral as well as octahedral site which may give partially inverse spinel ferrite characteristics. However, for sample annealed at 1000 °C, $Zn^{2+}$ ions in the octahedral site may have migrated to the tetrahedral site giving normal spinel characteristics. Also, the relative intensity ratio $I_{(hkl)}/I_{(400)}$ of the pure $Fe_3O_4$ cubic spinel reported [44] is compared to our cubic spinel to check the validation of the previous discussion. It was found that the relative intensity ratio of pure $Fe_3O_4$ comes which contain smaller $Fe^{3+}$ ions out to be less compare to the cubic spinel which contain larger radius $Zn^{2+}$ ions. These discussions can be justified more clearly with the analysis of Mössbauer Spectra.

A striking observation can be seen in Figure 1(b) that the splitting of (104) α-$Fe_2O_3$ planes and (311) cubic spinel phase with the increase of annealing temperature. These may be attributed to the reduction of crystal symmetry in a fraction of the respective phases where a symmetry equivalent planes splits into two different crystal planes which are very close to each other [48]. This reduction of the crystal symmetry could be related to the point defects (associated with the oxygen vacancies or Zn interstitial) present in the samples. Further, the Hopping length, also refer to as jump length, between the ions in the tetrahedral A-sites can be estimated by $L_A = a\frac{\sqrt{3}}{4}$ and in the octahedral sites by $L_B = a\frac{\sqrt{2}}{4}$ [46,47]. The plot of the Hopping length against the increase in annealing temperature of the samples is shown in Figure 2(d) indiactes that $L_A$ and $L_B$ increase with increasing annealing temperature which can be correlated with the cation size variation in the tetrahedral and octahedral sites.

### B. Photoluminescence analysis

Photoluminescence (PL) characteristics were studied for the as-synthesized and annealed sample to investigate the cation- and anion-induced defect states in the lattice. Figure 3(a) shows the evolution of PL spectra in ZF-W, ZF-W(300), ZF-W(500) and ZF-W(1000). The three emission bands found in the UV region 329 nm (~ 3.76 eV), 349 nm (~3.55 eV) and 390 nm (~3.17 eV) belong to the near band edge (NBE) emission of ZnO. These band are associated with the excitonic transition of electron-hole pair recombination. Basically, the PL emission spectra of Zn-site defects of ZnO is dominating over Fe based compound which shows fluorescence only in the nano-regime (size less than 20 nm), which can be seen from Figure 3(a) as well [49]. The broad defects related transition in visible region involves Zn interstitial ($Zn_i$), Zn-vacancy ($V_{Zn}$), oxygen interstitial ($O_i$), and oxygen vacancy ($V_o$). The blue emission at ~436nm (2.84 eV) is related to direct recombination of excited electron trapped in the $Zn_i$ defect states with the holes of valence band. It is also related with transition from conduction band to misplaced oxygen defects. Additionally, blue emission at ~468 nm (2.64 eV) is associated with electronic transition from $Zn_i$ to $V_{Zn}$ followed by non-radiative transition to valence band. The blue-green emission peak at ~491 nm (2.52 eV) emerges from the electronic transition from $Zn_i$ to $V_{Zn}$ or combine complex transition of $Zn_i$ and $V_o$ to valence band. The green emission peaks at ~512 nm (2.42 eV) is associated with $V_o$ or $V_{Zn}$ [50,51]. While the yellow-orange emission at ~550-620 nm (2.25-2.0 eV) is associated with neutral and charged $O_i$ defects (conduction band → $O_i$ defects) [52–54].

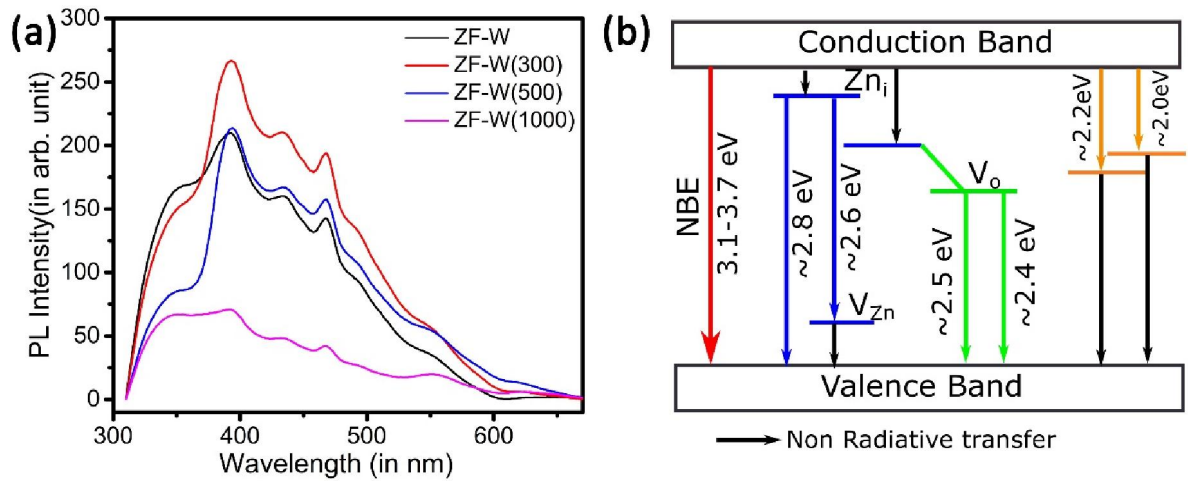

**Figure 3**: (a) PL spectra of ZFW, ZFW(300), ZFW(500) and ZFW(1000), (b) Schematics band diagram illustrating the emission band.

The PL emission in UV region here in the as-synthesized and annealed sample can be related with the concentration of oxygen in the samples [54]. Therefore, it is understood that the oxygen concentration in ZF-W(300) is higher compare to other samples. Annealing in the ambient atmosphere at low temperature, the presence of oxygen increases the stoichiometry of oxygen in the ZF-W(300). The lower PL intensity in ZF-W compare to ZF-W(300) in NBE region may be because of the presence of higher amount of non-radiative centre vacancy defects in ZF-W [55]. These vacancy defects are trapping the excited electrons that are supposed to recombine with the holes thereby decaying the PL emission non-radiatively. Also, there are charge vacancies that dispels the free electrons thereby decreasing carrier mobility [55]. These leads to longer time for electron–hole recombination rate thus a reduction in PL intensity. While in ZF-W(300), the density of non-radiative defects may have been very less and the emission is entirely due to radiative centres defects present on the surface and the lattice. While the concentration of oxygen in ZF-W(1000) is low at high annealing temperature due to thermal disorder resulting in quenching of PL intensity [56]. Also, there is a possibility of bonds breaking in some of the Zn-O linkage due to high thermal energy and thus generation of lattice site oxygen defects which could be the reason for observing peaks at around 616-624 nm region at ZF-W(500) and ZF-W(1000). For defects state region (at around 436-624 nm), the intensity of ZF-W(1000) decreases gradually (Figure 3(a)). This may be because of high rate of quenching of defects at higher temperature [52]. However, that is not the case for ZF-W(300) and ZF-W(500). With increase in annealing temperature, the ionization of neutral defects starts increasing up to certain activation energy temperature. The ionic lattice of ZF-W create an electrical potential field through which the neutral and ionized defects must pass in order to diffuse to the surface [52]. Ionized defects have a stronger interaction with this field gradient than the neutral counterpart. Therefore, at moderate annealing temperature, the energy accessible is comparable to activation energy barrier for the diffusion of neutral defects yet it is lower than the charge defects [52]. Hence, at such moderate temperature of 300 °C and 500 °C, most of the neutral defects are stabilized by ionization and thus the density of defects may be not be quenched. However, at a high annealing temperature of 1000 °C, this phenomenon does not occur and the existing defects in the course of plasma treatment were reduced. This explains the sharp decrease in the intensity of the PL emission spectra for the composite annealed at high temperature. However, new defects due to the migration of Zn from ZnO to

α-Fe$_2$O$_3$ might have occurred which is understood from the structural transformation of α-Fe$_2$O$_3$ to cubic spinel ferrites. A detailed schematics band diagram for the propose model is highlighted in Figure 3(b). A slight shifting of peaks centers is related to the variation in the width of the defects level [52]. Smaller the width in defect level, higher the average energy gap between the defects level and conduction band. As a result, a shift towards higher energy generating a blue shift is observed. Similar conclusion can be drawn for higher width which give rise to a red shift in the spectra.

### C. Mössbauer analysis

While PL analysis could predict the defects and the energy levels introduced in the ZnO, it gives least information about the cubic spinel ferrites and the hematite. The defects introduced in the Fe-containing compounds in the course of plasma treatment and the consequences after the heat treatment at various temperature can be explained by the Mössbauer spectroscopy. Mössbauer analysis of the all-studied sample at room temperature (298 K) shows the presence of α-Fe$_2$O$_3$, Fe$_3$O$_4$ and ZnFe$_2$O$_4$ in ZF-W (Table 3). The estimated relative amount of individual components obtained from the Mössbauer analysis is in good agreement with the XRD results (Table 1). The evolution of cubic spinel ferrites from the precursors (α-Fe$_2$O$_3$ and ZnFe$_2$O$_4$) was understood as a consequence to the exposure at high density, highly stable and homogeneous plasma jets [39]. The defect state developed during the exposure were quenched when deposited to the wall of the chamber. It is noteworthy to mention that the chamber wall was coupled with a chiller to maintain the temperature less than 10 °C. When the precursors were exposed under high-energy plasma, the cations also migrated along with the creation of oxygen vacancies during the exposure to a temperature of 2000 °C (as measured by spectroscopy method). The combinational motion of the oxygen vacancies and the cations modifies the structures and transform to the lowest energy structure. So, there is reduction of Fe$^{3+}$ to Fe$^{2+}$ and thus the transformation of α-Fe$_2$O$_3$ into cubic spinels (Fe$_3$O$_4$ and ZnFe$_2$O$_4$) occurs. Thus, it can be said that the ZF-W contains huge cationic and anionic point defects as well as the interfaces among the cubic spinel, α-Fe$_2$O$_3$ and ZnO. Subjecting the cations to the 300 °C, 500 °C and 1000 °C may enables the cations to reorganise from the structural lattices deformation, and the oxygen stoichiometry may reach towards the ideal condition. The ZF-W given heat treatment at 300 °C, the composition of the ZnFe$_2$O$_4$ remains same as that of ZF-W (36 %). While the quadrupole splitting (Δ) of the sextet increases to 0.19 mm/s, this is plausible only due to the structural transformation of magnetite to hematite. This is in agreement with the XRD, where the crystallite size at 300 °C of cubic spinel decreases as it is contributed only from the ZnFe$_2$O$_4$. The ZF-W(500) consist of a sextet having Δ value of 0.14 mm/s and a paramagnetic doublet (72 %). The hematite consists of 30 % of the total Fe containing compounds as calculated from the XRD and the sextet is found to have 28 %. Therefore, the sextet is understood to be contributed from α-Fe$_2$O$_3$. Interestingly the Δ value is low compared to the ideal synthetic α-Fe$_2$O$_3$ (~ 0.19-0.23 mm/s) [57–60]. The decrease in the Δ value may be due to the introduction of the Zn cations in the α-Fe$_2$O$_3$ lattice thereby increasing the degree of symmetry. From Figure 4(b), it is observed that the mean quadrupole splitting (<Δ>) of the sextet decreases with increase in the annealing temperature. The decrease in the <Δ> is due to the transformation of an asymmetric to symmetric structure due to the migration of Zn cations into the lattices of magnetite and α-Fe$_2$O$_3$. The <Δ> of the paramagnetic doublet increases at 300 °C which may be contributed to the structural transformation and obtained the ideal ZnFe$_2$O$_4$ structure at 500 °C as the Δ value is found to be

0.35 mm/s. Further increase in temperature at 1000 °C continues the migration of Zn into the lattices of α-Fe$_2$O$_3$ transforming to ZnFe$_2$O$_4$ with defects depicted by the increase in the Δ to 0.41 mm/s.

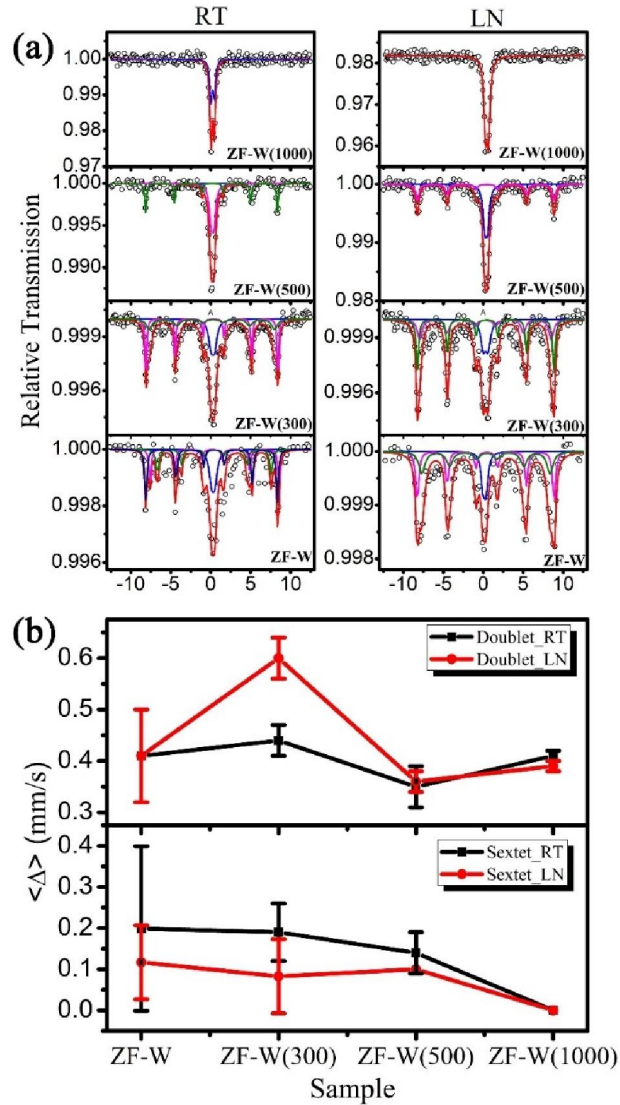

**Figure 4:** (a) Mossbauer spectra of the samples ZF-W, ZF-W(300), ZF-W(500) and ZF-W(1000) at RT and LN, RT means Room Temperature and LN means Liquid Nitrogen. (b) Variation in average quadrupole splitting (<Δ>) of paramagnetic sextet and doublet of the sample annealed at various temperature.

To explore more on the annealing effect of the samples, Mössbauer spectra were taken at -190 °C using liquid nitrogen. The increase in Δ (0.60 mm/s) values for paramagnetic doublet for ZF-W(300) compare to the values of ZF-W is due to the presence of defects [61]. It shows that ZnFe$_2$O$_4$ phase in ZF-W(300) is in a metastable state where defects are present. However, the Δ value of ZF-W(500) decreases to 0.36 mm/s which is very similar to the reported value of synthetic ideal ZnFe$_2$O$_4$. This shows that ZnFe$_2$O$_4$ in ZF-W(500), although phase transformation occurs, attains an equilibrium point where there is no involvement of cation migration or substitution. Therefore, the contribution of vacancy defects from this state is negligible. Further, the Δ value of ZF-W(1000) again increases to 0.39 mm/s. This could be explained as due to the introduction of oxygen vacancy develop as a consequence of the migration of $Zn^{2+}$ cation to the $Fe^{3+}$ lattice site completely transforming α-$Fe_2O_3$ to $ZnFe_2O_4$ at 1000°C. Additionally, the oxidation of the cubic spinel $Fe_3O_4$ to α-$Fe_2O_3$ occur with the increase in the annealing temperature on the boundary of the cubic spinel. This can be validated with the increase of α-$Fe_2O_3$ content from 31% for ZF-W to 44% for ZF-W(500). Moreover, there is decrease in the <Δ> of sextets which can be observed in Figure 4(b). This also indicates the variation in the neighbouring atoms of Fe caused by the emergence of defect states nearby it.

Table 2 Mossbauer analysis data for the prepared samples.

| | | Isomer shift | Quadruple splitting | Hyperfine splitting | Area (%) | Phase |
|---|---|---|---|---|---|---|
| Sample tested at 25 °C | ZF-W | 0.41(3) | 0.41(0) | - | 36 | $ZnFe_2O_4$ |
| | | 0.37(2) | 0.32(5) | 51.1(0) | 27 | $Fe_2O_3$ |
| | | 0.42(3) | 0.26(8) | 50.0(4) | 15 | $Fe_A$ |
| | | 0.54(4) | 0.009(7) | 44.2(3) | 22 | $Fe_B$ |
| | ZF-W (300) | 0.37(1) | 0.44(3) | - | 36 | $ZnFe_2O_4$ |
| | | 0.35(1) | -0.19(2) | 51.2(7) | 19 | α-$Fe_2O_3$ |
| | | 0.33(2) | -0.19(5) | 48.7(2) | 45 | Intermediate of α-$Fe_2O_3$ and $Fe_3O_4$ |
| | ZF-W (500) | 0.33(2) | 0.35(4) | - | 72 | $ZnFe_2O_4$ |
| | | 0.27(3) | -0.14(5) | 51.2(9) | 28 | α-$Fe_2O_3$ |
| | ZF-W (1000) | 0.34(1) | 0.41(1) | - | 100 | $ZnFe_2O_4$ |
| Sample tested at -190 °C | ZF-W | 0.28(5) | 0.41(9) | - | 15 | $ZnFe_2O_4$ |
| | | 0.47(2) | -0.13(0) | 53.2(2) | 31 | α-$Fe_2O_3$ |
| | | 0.48(5) | -0.11(9) | 49.8(2) | 54 | Intermediate of α-$Fe_2O_3$ and $Fe_3O_4$ |
| | ZF-W (300) | 0.40(2) | 0.60(4) | - | 29 | $ZnFe_2O_4$ |
| | | 0.36(3) | 0.02(6) | 51.1(2) | 41 | Cubic IONP |
| | | 0.48(2) | -0.17(3) | 53.1(6) | 30 | α-$Fe_2O_3$ |
| | ZF-W (500) | 0.46(1) | 0.36(2) | - | 56 | $ZnFe_2O_4$ |
| | | 0.47(2) | -0.10(0) | 53.08(2) | 44 | α-$Fe_2O_3$ |
| | ZF-W (1000) | 0.47(1) | 0.39(1) | - | 100 | $ZnFe_2O_4$ |

### D. Propose Mechanism of phase transformation and defects evolution

Based on the above experimental observations, the phase transformation of ZF-W annealed at different temperature may be described according to the illustration in Figure 5. Cation substitution of $Zn^{2+}$ to $Fe^{3+}$ which transforms to $ZnFe_2O_4$ takes place in the course of heating at different temperature. It is evident from the increase in the relative amount of $ZnFe_2O_4$ for the samples annealed above 500 °C, as obtained from XRD and Mössbauer spectra. Annealing ZF-W at 300 °C oxidized the cubic iron oxides, forming a layer of α-$Fe_2O_3$ phase on the surface. This fact results to the decrease in the thickness (size) of the cubic spinel ferrites, as depicted from the XRD. At 500 °C, $ZnFe_2O_4$ attains an equilibrium state which is quite similar to the ideal properties of as-synthesized $ZnFe_2O_4$. Also, the cubic iron oxides are oxidized due to heating and transform to α-$Fe_2O_3$ forming a new layer of oxidize α-$Fe_2O_3$ in addition to the existing hematite in ZF-W. However, at 1000 °C, $Zn^{2+}$ enters the α-$Fe_2O_3$ lattice thereby completely transforming to defective $ZnFe_2O_4$ structure. Moreover, the defect states emerging from ZnO seems to increase as the annealing temperature increases until 1000 °C where it decreases. This is noticeable from the PL spectra, as discussed earlier. The defects states of cubic spinel as well as α-$Fe_2O_3$ was elucidated from the isomer shift and quadrupole splitting data analysed from the Mössbauer spectra. It is very crucial for ZF-W(1000) because oxygen vacancy defects may form when $Zn^{2+}$ cation lattice enters $Fe^{3+}$ to form $ZnFe_2O_4$. Emergence of oxygen vacancies defects when annealed at high temperature is already reported [61]. Thus, a new oxygen vacancy defect states may have been created for ZF-W(1000) sample, which was discussed in Mössbauer analysis.

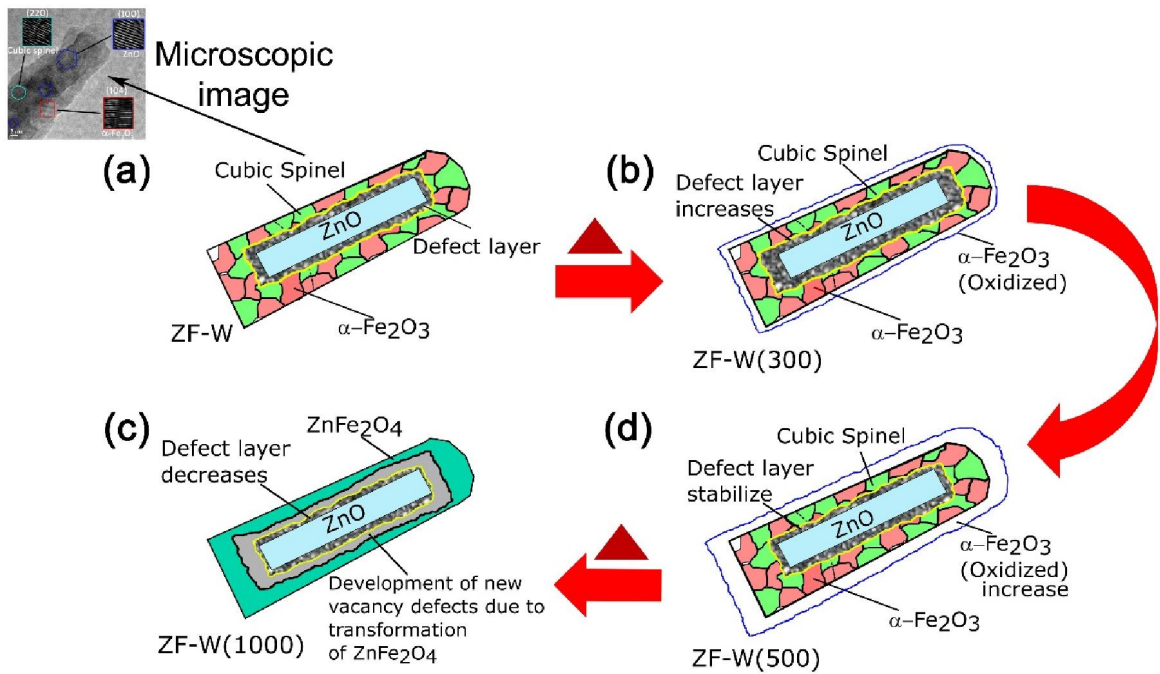

**Figure 5:** Schematic diagram of structural transformation and changes in the defects for (a) ZF-W, (b) ZF-W(300), (c) ZF-W(500) and ZF-W(1000)

### E. Methyl blue dye-adsorption properties

The catalytic activity of the all studied sample were tested against the MB dye under dark condition (Figure 6(a)). it is evident that the degradation rate of ZF-W is comparable with the annealed ZF-W(300) and ZF-W(1000). Specifically, the estimated MB degradation efficiency for the ZF-W, ZF-W(300), ZF-W(500) and ZF-W(1000) are 86%, 84% 68% and 82%, respectively. A clear trend can be seen from the adsorption result that the degradation efficiency is highest for ZF-W, decreases with the increasing annealing temperature, lowers at 500 °C and recovers at 1000 °C. It was reported [39] that the degradation of MB dye is related with the charge-transfer capability and surface defects (mainly $V_o$ and $V_{Zn}$). The vacancies provides active sites for fast redox reaction and creates localized level below the conduction band which helps in lowering the bandgap [62]. This leads to faster redox reaction and faster degradation of the MB dye. The fast degradation of ZF-W, ZF-W(300) and ZF-W(1000) can be explained by the concentration of various defects present in them as discussed earlier in the PL analysis. However, despite the presence of defect states in ZF-W(500), it shows less degradation of MB dye compared to other samples. This is because of the fact that oxygen vacancy defects present in the sample due to contribution from ZnFe$_2$O$_4$ structure is negligible, as discuss in the Mossbauer analysis. So, the overall vacancy defects in ZF-W(500) may have been less compare to ZF-W(300) and ZF-W(1000) where oxygen vacancy defect from the ZnFe$_2$O$_4$ is present. So, defects arises in ZF-W(1000) due to ZnO may have less effect and the higher degradation efficiency in ZF-W(1000) may be due to the evolution of vacancies arising as a consequence of phase transformation. To support the preceding discussion, pure ZnO was annealed at 1000 °C and the effect of MB dye degradation was examined. It was found that ZnO annealed at 1000 °C shows very less activity with a degradation efficiency of only ~9%. This clarifies that degradation of MB dye on ZF-W(1000) is entirely related with the defects evolve due to the complete phase transformation of ZnFe$_2$O$_4$.

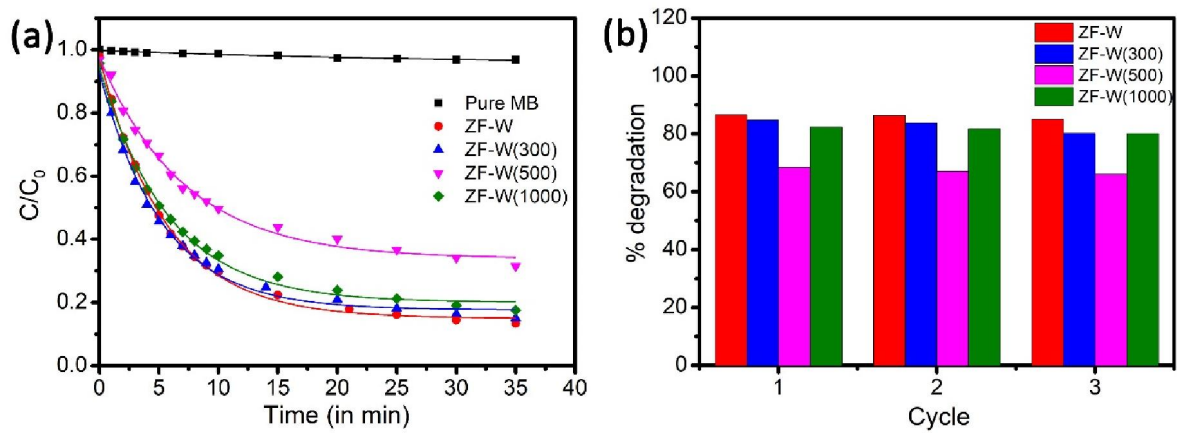

**Figure 5**: (a) Degradation of MB dye for ZF-W, ZF-W(300), ZF-W(500) and ZF-W(1000), (b) reusability of the samples for 3 cycles.

The possible mechanism for the degradation of MB dye may be the evolution of $^{\bullet}$OH radical and $O_2^{\bullet-}$ radical when MB dye reacts with the catalyst. The adsorption of MB dye on to the catalyst could not be completely neglected although no interaction with sulfonic group of the MB dye with catalyst was found in the previous work [39]. Thus, major contribution for the decoloration of MB dye directs towards the surface defects. The mechanism underlying this degradation process can be described in terms of the extent of oxygen vacancies in the sample. The primary step is the dissociation of water molecules at oxygen-vacant sites. When precipitates come into contact with water molecules (H$_2$O), it splits into hydrogen (H$^{\bullet}$) radical and hydroxyl ($^{\bullet}$OH) radical due to the interaction with oxygen vacancies and oxygen atom at the surface of the catalyst as shown in Eq. (1) [63,64]. The dissociated H$^{\bullet}$ radical forms another $^{\bullet}$OH radical and oxygen (O$^{\bullet}$) radical with the surface oxygen atom in Eq. (2). The new O$^{\bullet}$ radical then interacts with other H$_2$O molecules to form two more $^{\bullet}$OH radical shown in Eq. (3). These $^{\bullet}$OH radical provides oxidation of the MB dye and thus the degradation occurs.

$$H_2O + O_{vacancy} \rightarrow H^{\bullet} + HO^{\bullet} \tag{1}$$

$$H^{\bullet} + O_2 \rightarrow HO^{\bullet} + O^{\bullet} \tag{2}$$

$$O^{\bullet} + H_2O \rightarrow 2HO^{\bullet} \tag{3}$$

$$Dye + HO^{\bullet} \rightarrow Degradation \tag{4}$$

Moreover, MB dye interacts on the surface of ZF-W where a redox reaction is instigated with the formation of electron-hole pair (exciton). However, as discussed earlier in PL there are some free-electrons trapped in the vacancy or dispel due to charge defects which prevents from electron–hole recombination. Thus, the free electron from conduction band give rise to oxygenated free radicals such as OH• through the reduction of $O_2$ (Eq. 5-9) [65,66]. Holes in the valence band also provide HO• radical by the ionization and oxidation of water molecules [65]. The evolve OH• are strong oxidant which oxidize the MB dye.

$$O_2 + e^{-}_{CB} \rightarrow O_2^{•-} \tag{5}$$

$$O_2^{•-} + H_2O \rightarrow HO_2^{•} + HO^{-} \tag{6}$$

$$HO_2^{•} + HO_2^{•} \rightarrow H_2O_2 + O_2 \tag{7}$$

$$H_2O_2 + O_2^{•-} \rightarrow HO^{•} + HO^{-} + O_2 \tag{8}$$

$$H_2O + ZFW(h^{+}_{VB}) \rightarrow ZFW + HO^{•} \tag{9}$$

Further, reusability and stability of the as synthesized and annealed samples against the MB dye were examined. From Figure 6(b), it is found that the regenerative ability of the samples was quite impressive. ZF-W shows more stability than the annealed samples in terms of the degradation of MB dye. From this study, the defect-rich ZF-W with admirable constituents (ZnO, α-$Fe_2O_3$ and cubic spinel ferrites) is identified as a most suitable one for the MB dye degradation.

## IV. Conclusion

A composite of ZnO and α-$Fe_2O_3$ has been synthesized at the plasma-deposition wall by exposing the precursor at high-energy plasma. Complex interaction of structural phase transformation of the composites was studied by annealing them at various temperature of 300, 500 and 1000 °C. It has been shown that, the lattice parameter and crystallite size increases with the annealing temperature with the exception of sample annealed at 300 °C. Also, substitution of a smaller $Fe^{3+}$ ion with larger $Zn^{2+}$ ions increase the relative intensity ratio of the tetrahedral site. The structural phase transformation of $ZnFe_2O_4$, oxidation of $Fe_3O_4$ to α-$Fe_2O_3$ with defects were observed while annealing at 300 and 500 °C. The emergence of defects from ZnO were detailed using PL and the defects related to α-$Fe_2O_3$ and cubic spinel were established using Mössbauer spectroscopy. Vacancy defects acts as a quencher for the emission of spectra. Compared to other, ZF-W(500) have a stable equilibrium structure where defects are equilibrated depicted by the quadrupole splitting value comparable to ideal synthetic $ZnFe_2O_4$. At 1000 °C, a new layer of vacancy defect states are formed as a result of migration of Zn cation into Fe sites of α-$Fe_2O_3$. Further, degradation of MB dye on the annealed samples were found to complete under 30 min without any external energy source. ZF-W(300) and ZF-W(1000) shows a comparable degradation efficiency of 84% and 82% respectively with that of ZF-W (~86%). However, the sample ZF-W(500) shows a low degradation efficiency of 68% only which could be related with the stable low defect state of $ZnFe_2O_4$. The difference in temperature cause anomalies in the cation arrangement and composition of the samples led by the evolution of various defect states. This study provides a better understanding of the thermal annealing effect and its role in defects establishment carrying forward.

## V. ACKNOWLEDGEMENTS

The authors (B.W. and L.H. S.) thank the DST-SERB for the financial assistance under the project having file no. CRG/2021/001611. We would like to thank Department of Chemistry, NIT Manipur for XRD and PL. S. G thanks to European commission for Seal of Excellence award under the Horizon 2020's Marie Skłodowska-Curie actions.

**Declaration**

There are no conflict to declare

**Conflict of Interest**

There are no conflict of interest